\documentclass[10pt,twocolumn,letterpaper]{article}

\usepackage{iccv}
\usepackage{times}
\usepackage{epsfig}
\usepackage{graphicx}
\usepackage{subcaption}
\usepackage{amsmath}
\usepackage{pifont}
\newcommand{\cmark}{\text{\ding{51}}}
\newcommand{\xmark}{\text{\ding{55}}}
\usepackage{amssymb}
\usepackage{adjustbox}
\usepackage{authblk}

 \iccvfinalcopy % *** Uncomment this line for the final submission

\ificcvfinal\fi
\begin{document}

%TODO: 
%   1. ADD REFERENCES TO ROBOT NEVIGATION...

%ideas for supplementry meterial:
%  1. detailed architectures
%  2. surfaces that can be achieved from one layer
%  3. scene segmentation
%  4. failure cases
%%%%%%%%% TITLE
\title{Momen$^e$t: Flavor the Moments 
 in Learning to Classify Shapes}

\author[1, 2]{Mor Joseph-Rivlin}
\author[1]{Alon Zvirin}
\author[1]{Ron Kimmel}

\affil[1]{Technion - Israel Institute Of Technology}
\affil[2]{ Rafael Ltd, Israel}
\affil[ ]{\tt\small {\{mor1joseph@campus,salz@cs,ron@cs\}.technion.ac.il}}
%\author{Mor Joseph-Rivlin\\
%EE Dept. Technion IIT\\
%Haifa 32000, Israel\\
%{\tt\small mor1joseph@campus.technion.ac.il}
%\and
%Alon Zvirin\\
%CS Dept. Technion IIT\\
%Haifa 32000, Israel\\
%\and
%Ron Kimmel\\
%CS Dept. Technion IIT\\
%Haifa 32000, Israel\\
%}
%\author{Mor Joseph-Rivlin$^1$ \and Alon Zvirin$^2$ \and Ron Kimmel$^2$}
%\date{%
%    $^1$ Technion EE department, Haifa, Israel\\%
%    $^2$ Technion CS department, Haifa, Israel\\[2ex]%
%}

\maketitle
%\thispagestyle{empty}

%%%%%%%%% ABSTRACT
\begin{abstract}
A fundamental question in learning to classify 3D shapes is how to treat the data in a way that would allow us to construct efficient and accurate geometric processing and analysis procedures. 
Here, we restrict ourselves to networks that operate on point clouds. 
There were several attempts to treat point clouds as non-structured data sets by which a neural network is trained to extract discriminative properties.

The idea of using 3D coordinates as class identifiers motivated us to extend this line of thought to that of shape classification by comparing attributes that could easily account for the shape moments. 
Here, we propose to add polynomial functions of the coordinates allowing the network to account for higher order moments of a given shape. 
Experiments on two benchmarks show that the suggested network is able to provide state of the art results and at the same token learn more efficiently in terms of memory and computational complexity. 
\end{abstract}

%%%%%%%%% BODY TEXT
\section{Introduction}

In recent years the popularity and demand for 3D sensors has vastly increased. 
Applications using 3D sensors include robot navigation, stereo vision, and advanced driver assistance systems to name just a few.
Recent studies attempt to adjust deep neural networks (DNN) to operate on 3D data representations for diverse geometric tasks.
Motivated mostly by memory efficiency, our choice of 3D data representation is coordinates of point clouds. 
One school of thought indeed promoted feeding these geometric features as input to deep neural networks that operate on point clouds for classification of rigid objects. 

From a geometry processing point of view, it is well known that moments characterize a surface and can be useful for the classification task.
To highlight the importance of moments as class identifiers, we first consider the case of a continuous surface. 
In this case, geometric moments uniquely characterize an object.
Furthermore, a finite set of moments is often sufficient as a compact signature that defines the surface  \cite{bronstein2008numerical}. 
This idea was classically used for estimation of surface similarity. 
For example, if all moments of two surfaces are the same, the surfaces are considered to be  identical.
Moreover, sampled surfaces, such as point clouds, can be identified by their estimated geometric moments, where it can be shown that the error introduced by the sampling is proportional to the sampling radius and uniformity. 
\begin{figure}[htbp]
	\centering
	\centerline{\includegraphics[width=1\columnwidth]{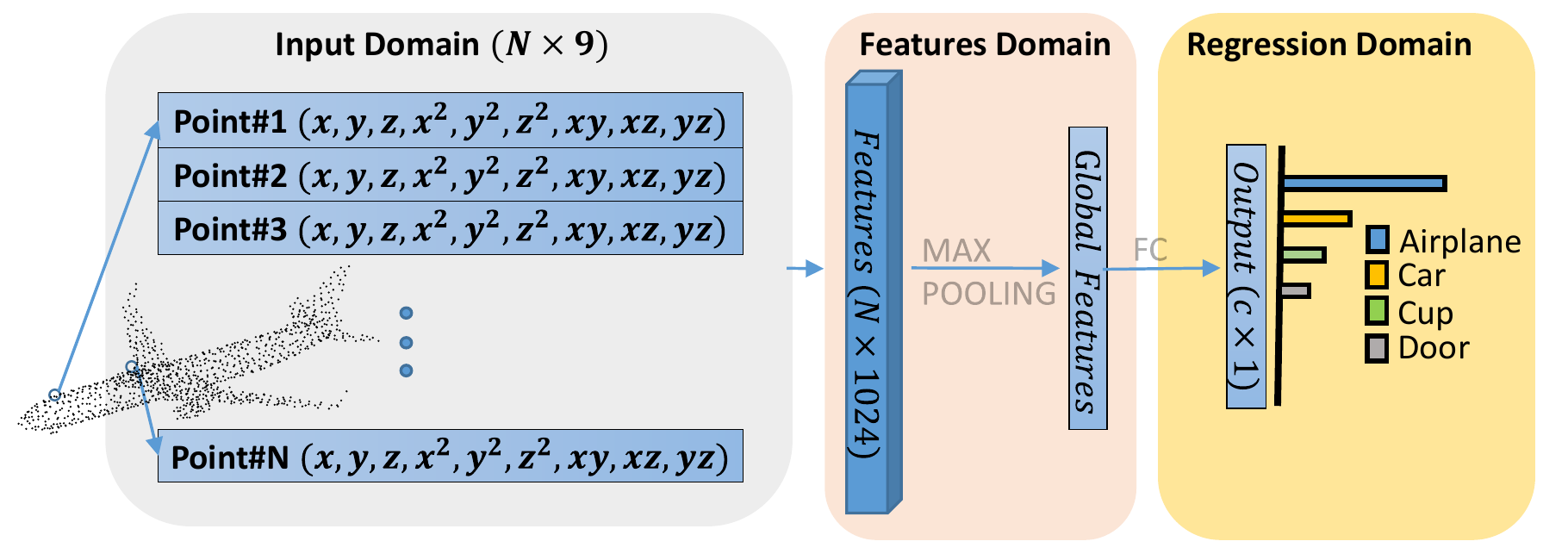}}
	\caption{Illustration of the proposed object classification architecture. 
	The input of the network includes the point cloud coordinates as well as second order polynomial functions of these coordinates. 
    It enables the network to efficiently learn the shape moments.} 
    %Fc denotes a fully connected layer.}
	\label{fig:archSmall}
\end{figure}
Our goal is to allow a neural network to simply lock onto variations of geometric moments. 
One of the main challenges of this approach is that training a neural network to approximate polynomial functions requires the network depth and complexity to be logarithmically inversely proportional to the approximation error \cite{yarotsky2017error}.
In practice, in order to approximate polynomial functions of the coordinates for the calculation of geometric moments the network requires a large number of weights and layers. 
Qi et al. \cite{qi2017pointnet} proposed a network architecture which processes point clouds for object classification.  
The framework they suggested includes lifting the coordinates of each point into a high dimensional learned space, while ignoring the geometric structure. An additional pre-processing transformation network was supposed to canonize a given point cloud, yet it was somewhat surprising to discover that the transformer results are not invariant to the given orientations of the point cloud.
Learning to lift into polynomial spaces would have been a challenge using the architecture suggested in \cite{qi2017pointnet}.
At the other end, networks that attempt to process other representations of low dimensional geometric structures such as meshes, voxels (volumetric grids), and multi-view projections are often less efficient when considering both computational and memory complexities.

In this paper, we propose a new network that favors geometric moments for point cloud object classification. 
The most prominent element of the network is supplementing the given point cloud coordinates  together with polynomial functions of the coordinates, see Fig.\ref{fig:archSmall}.
This simple operation allows the network to account for higher order moments of a given shape. 
%The proposed network implementation is based on a simplified version of the pointNet architecture, with only one layer in the feature domain. 
Thereby, the suggested network requires relatively low computational resources in terms of run-time, and memory in sense of the number of network's parameters. 
Experiments on two benchmarks show that the suggested scheme is able to learn more efficiently compared to previous methods in terms of memory and actual computational complexity while providing more accurate results. %
Lastly, it is easy to implement the proposed concept by just calculating the polynomial functions and concatenating them as an additional vector to the current input of point cloud coordinates.

\section{Related Efforts}

This section reviews some early results relevant to our discussion. 
First, we relate to methods used for learning from processed point clouds such as voxels and trees based models.
Next, we present a line of works that consume directly point clouds for object classification.
%AZ: "similarly to our method" (NOT BEST TIME TO REOPEN A DISCUSSION...)
%AZ: but as it is now, you do process the point-clouds (the kNN)
%AZ: can keep AS IS but careful about "In contrast to the methods mentioned above which encode point clouds in trees or in voxels, our method consumes the points directly."  appearing later
The third part describes early studies of {\it higher order networks}, in which each layer applies polynomial functions to its inputs, defined by the previous layer's output. 
We provide evidence that similar simple lifting ideas were applied quite successfully to geometric object recognition and classification in the late 80's.
\newline

\subsection{Learning features from processed point clouds}  %encoded
%AZ: I think the title is OK.
%AZ: maybe just "processed", because some of the methods you describe do pre-processing, others do the processing as part of learning features.  MJ:FIXED
The most straightforward way to apply convolutional neural networks (CNNs) to 3D data is by transforming 3D models to grids of voxels, see for example \cite{maturana2015voxnet,wu20153d}. 
A grid of occupancy voxels is produced and used as input to a 3D CNN.
This approach has produced successful results, but has some disadvantages such as loss of spatial resolution, and the use of excessively large memory.
For some geometric tasks that require analysis of fine details, in some cases, implicit (voxel) representation would probably fail to capture fine features.
Several methods replace the grid occupancy representation with radial basis functions \cite{Atzmon:2018:PCN:3197517.3201301}, fisher vectors \cite{ben20183dmfv} and mean points \cite{hua2018pointwise} for convolving with 3D kernels. However, the 3D grid representation has an inherent drawback in terms of memory consumption.

KdNet \cite{klokov2017escape} and OcNet \cite{riegler2017octnet} approaches exploit tree models to build a balanced and unbalanced (respectively) hierarchical structure partitions as a prepossessing stage. Nevertheless, rotation, noise or variation in number of points force rebuilding those trees from scratch.

In contrast to the methods mentioned above which encode point clouds in trees or in voxels, our method consumes the points directly.
\newline

\subsection{Learning features directly from point clouds} 
A deep neural network applied to point clouds known as {\it pointNet}  was introduced in \cite{qi2017pointnet}.
That architecture processes the points' coordinates for classification and segmentation.
The classification architecture is based on fully connected layers and symmetry functions, like max pooling, to  establish invariance to potential permutations of the points. 
In addition, all Multi-Layer Perceptrons (MLPs) operations are performed per point, thus, interrelations between points are accomplished only by weight sharing.
Furthermore, the pointNet architecture also contains a transformer network for coping with input transformations. It is supposed to learn a set of transformations that transform the geometric input structure into some canonical configuration, however it is computationally expensive.
The architecture pipeline commences with MLPs to generate a per point feature vector, then, applies max pooling to generate global features that serve as a signature of the point cloud. 
Finally, fully connected layers produce output scores for each class. 

PointNet's main drawback is its limited ability to capture local structures which has lead to an extensive line of work. 
PointNet++ \cite{qi2017pointnet++} extracts features in local multiscale regions and aggregates local features in hierarchical manner. 
RSNet \cite{huang2018recurrent} employs recurrent neural networks (RNN) on point clouds slices for features extraction. 
KC-Net \cite{shen2018mining} uses point cloud local structures by kernel correlation and graph pooling. 
DGCNN \cite{wang2018dynamic} builds a k-nearest neighbor (kNN) graph in both point and feature spaces to leverage neighborhood structures. 
SO-Net \cite{li2018so} reorganizes the point cloud into a 2D map and learns node-wise features for the map. 
Those methods indeed capture the local structures by explicit or handcrafted methods of encoding the local information and have achieved state of the art results in 3D shape classification and segmentation tasks. 
However, the use of geometry context in the form of geometric moments of 3D shapes is still absent from the literature.
\newline 

%___________________________________________________
\subsection{Learning high order features} 
\label{HONN} 
{\em Multi-layer perceptron} (MLP) is a neural network with one or more hidden layers of perceptron units. 
The output $\phi$ of such a unit with an activation function $\sigma$, previous layer's output $\eta$ and vector of learned weights $w$ is a first order perceptron, defined as $\phi=\sigma(\sum_{j} w_j \eta_j)$.
Where, $\sigma$ is a sigmoid function, $\sigma(a) =1/(1+e^{-a})$.

In the late 80's, the early years of artificial intelligence, Giles et al. \cite{giles1987learning,giles1988encoding} proposed extended MLP networks called {\em higher-order neural networks}.
Their idea was to extend all the perceptron units in the network to include also the sum of products between elements of the previous layer's output $\eta$.
The extended perceptron unit named high-order unit is defined as 

\begin{eqnarray} \label{eu_honn}
\resizebox{.9 \linewidth}{!} 
{
$\phi=\sigma\left(\sum_{i}w_{i}\eta_i+\sum_{i,j} w_{ij} \eta_i  \eta_j +\sum_{i,j,k} w_{ijk} \eta_i \eta_j \eta_k +\ldots\right)
$}
\end{eqnarray}
%AZ: don't understand "Missing $ inserted" error
These networks included some or all of the summation terms. 
Theoretically, an infinite term of single high order layer can perform any computation of a first order multi-layer network \cite{lee1986machine}.
Moreover, the convergence rate using a single high layer network is higher, usually by orders of magnitude, compared to the convergence rate of a multi-layer first order network \cite{giles1988encoding}. 
Therefore, higher-order networks are considered to be powerful, yet, at the cost of high memory complexity. 
The number of weights grow exponentially with the number of inputs, which is a prohibitive factor in many applications.

A special case of high order networks is the {\it square multi-layer perceptron} proposed by Flake et al. \cite{flake1998square}.
They extend the perceptron unit with only the squared components of $\eta$, 
 given by % see eq. \ref{eu_honn2}.
\begin{eqnarray} \label{eu_honn2}
 \phi&=&\sigma\left (\sum_{i} w_i \eta_i +\sum_{j} w_{j} \eta_j^2 \right).
\end{eqnarray}
The authors have shown that with a single hidden unit the network has the ability to generate localized features in addition to spanning large volumes of the input space, while avoiding large memory requirements.

%=================================================================================================================

\section{Methods}

The main contribution of this paper is leveraging the network's ability to operate on point clouds by adding polynomial functions to their coordinates. 
Such a design can allow the network to account for higher order moments and therefore achieve higher classification accuracy with lower time and memory consumption.
Next, we show that it is indeed essential to add polynomial functions to the input, as learning to multiply inputs items is a challenge for neural networks.
\newline

\subsection{Problem definition}
The goal of our network is to classify 3D objects given as point clouds embedded in $\mathbb{R}^3$. 
A given point cloud $X$ is defined as a cloud of $n$ points, where each point is described by its coordinates in $\mathbb{R}^3$. 
That is, $X = \{{\bf x}_1,\ldots,{\bf x}_n\}$, where
 each point ${\bf x}_j$, is given by its coordinates $(x_j,y_j,z_j)^T$.  
The output of the network, should allow us to select the class $Y \in \mathbf{L}$, where $ \mathbf{L} = \{1, \ldots , L\}$ is a set of $L$ labels.
For a neural network defined by the function $g:\mathbb{R}^{n\times 3} \rightarrow \mathbb{R}^L$, the desired output is a score vector in $\mathbb{R}^L$, such that $Y = \arg\max_{l\in \mathbf{L}} g(X;l)$.
\newline
%{\color{red}\\ Please check that the above is what you wanted to say...
%} Yes! thanks.

%____________________________________________
\subsection{Geometric Moments} 
\label{NNgeom} 
The usage of invariant moments as shape descriptors is based on the theory of invariant algebra \cite{grace1903algebra} which addresses mathematical objects that remain unchanged under linear transformations. Sadjadi and Hall \cite{sadjadi1980three}, were the first to utilize moments as 3D shape descriptors. 
Since the 80’s, extensive research has been done exploring moments as descriptors of shapes in two and three dimensions \cite{hu1962visual,abu1984recognitive,sadjadi1978numerical,teh1988image,khotanzad1990invariant,reeves1988three}.

Geometric moments of increasing order represent distinct spatial characteristics of the point cloud distribution, implying a strong support for construction of global shape descriptors.
By definition, first order moments represent the extrinsic centroid; second order moments measure the covariance and can also be thought of as moments of inertia. 
Second order moments of a set of points $X \subseteq \mathbb{R}^3$ can be compactly expressed in a $3\times 3$ symmetric matrix $\Sigma$, Eq. (\ref{eu_discreteMom}).   where ${\bf x}_j  \in X$ defines a point given as a vector of  its coordinates ${\bf x}_j = (x_j,y_j,z_j)^T$. 
\begin{equation} \label{eu_discreteMom}
 \Sigma = \sum_j {\bf x}_j{\bf x}_j^T.
\end{equation}

\begin{figure}[htbp]
	\centering
	\centerline{\includegraphics[width=0.6\columnwidth]{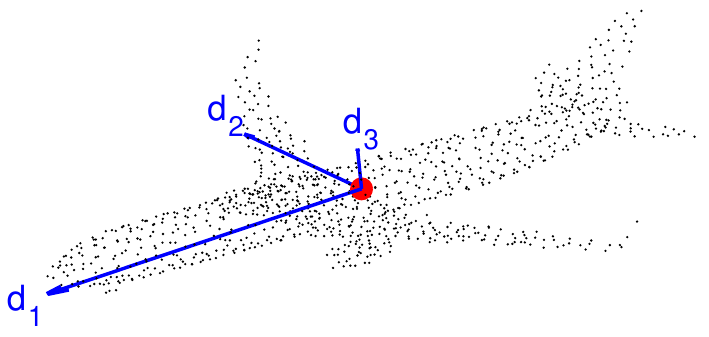}}
	\caption{The first and second geometric moments displayed on a point cloud. 
    Using the first order moments (red disc), the translation ambiguity can be removed. 
    The principal directions $d_1, d_2, d_3$ (blue arrows) are defined by the second order geometric moments. 
Adding these moments to the input helps the network to resolve the rotation ambiguity.}
	\label{fig:mom}
\end{figure}

\begin{figure*}[htbp]\label{x2}
    \centering
    \mbox{
        \includegraphics[width=0.4\textwidth,height=5cm]{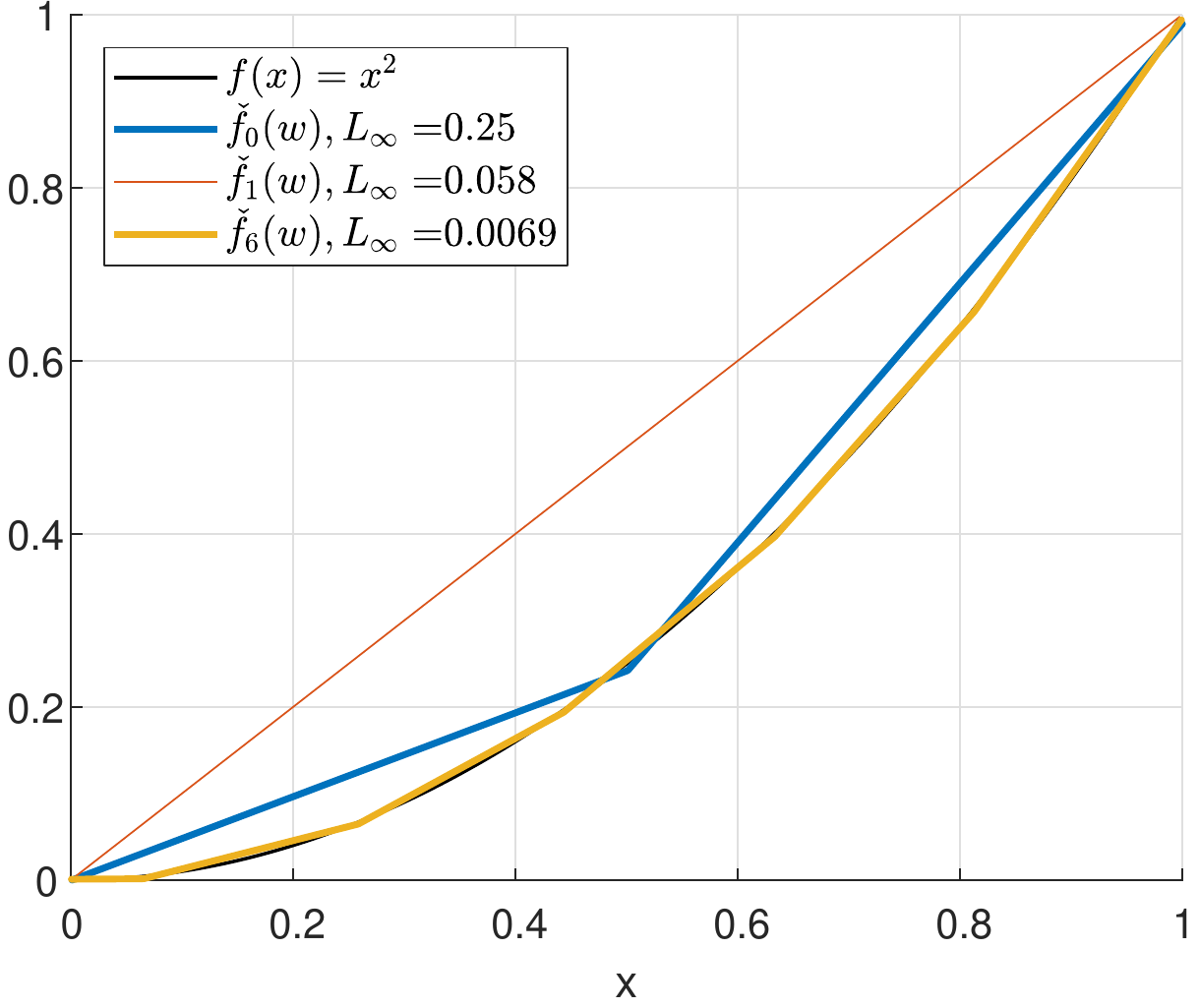} ~
        \includegraphics[width=0.4\textwidth,height=5cm]{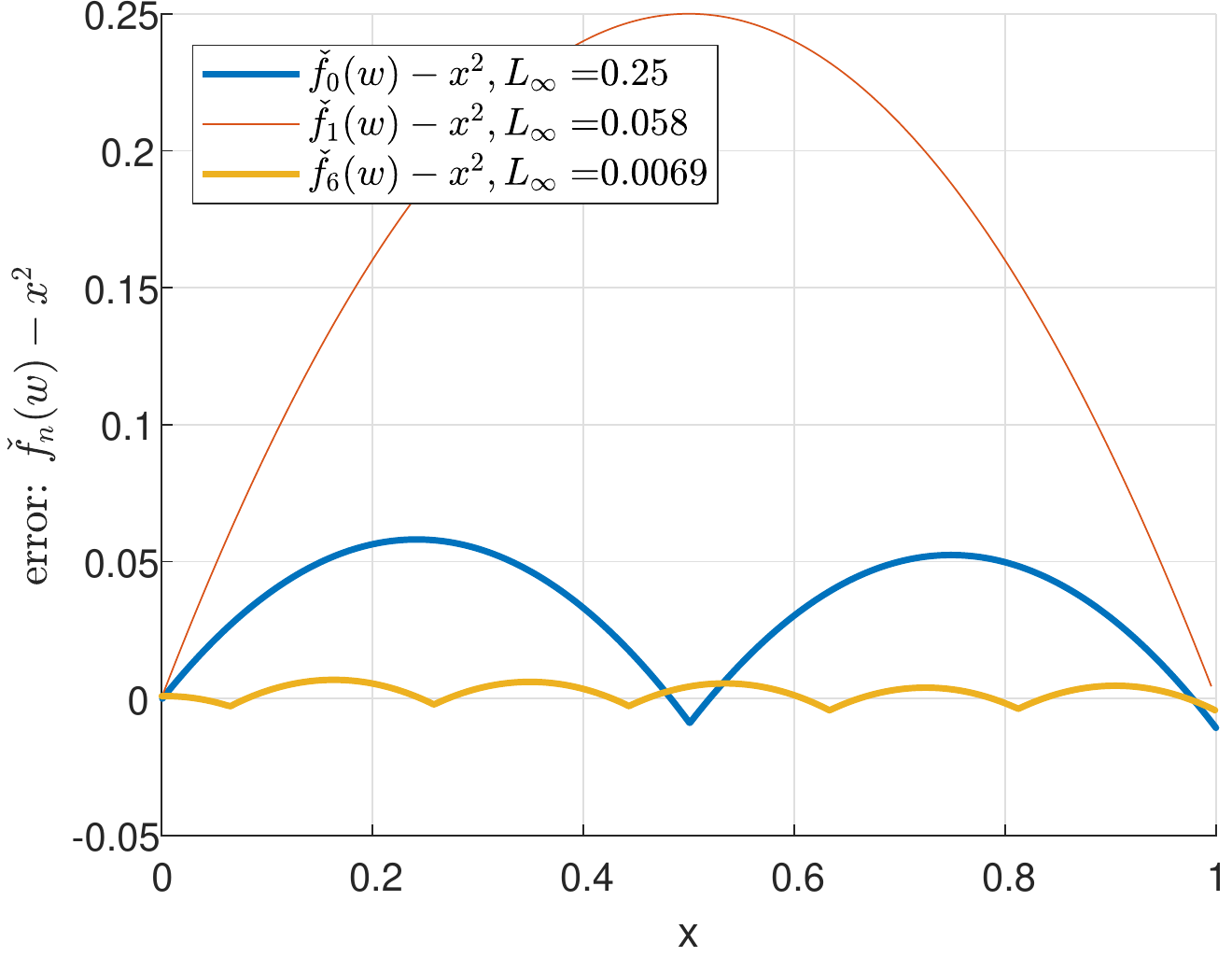}
    }
    %AZ: missing ylabel in left figure
    \caption{
    Approximation of the function $x^2$ by a fully-connected feed forward neural network $f_n(w)$ where $n$ denotes the number of hidden layers. 
    Left: Approximation by different numbers of hidden layers. 
    Right: The error of these functions.}
\label{fig:x2}
\end{figure*}

Roughly speaking, we propose to relate between point clouds by learning to implicitly correlate their moments.
Explicitly, the functions $(x^2,y^2,z^2,xy,xz,yz)$ of each point are given to a neural network as input features in order to obtain better accuracy.
\newline

\noindent \textit{Geometric transformations.} 
A desired geometric property is invariance to rigid transformations.
Any rigid transformation in $\mathbb{R}^3$ can be decomposed into rotation and translation transformations, each defined by three parameters \cite{bronstein2008numerical}. 
A rigid Euclidean transformation $T$ operating on a vector $v \in \mathbb{R}^3$ has the general form
\begin{equation} 
\label{eu_TRANS}
T(v) = R\cdot v + t, 
\end{equation}
where $R$ is the rotation matrix and $t$ is the translation vector.

Once translation and rotation are resolved, a canonical form can be realized.
The pre-processing procedure translates the origin to the center of mass given by the first order moments, and scales it into a unit sphere compensating for variations in scale. 
The rotation matrix, determined by three degrees of freedom, can be estimated by finding the principal directions of a given object, for example see Figure \ref{fig:mom}.

\begin{figure*}[htbp]
	\centering
	\centerline{\includegraphics[width=1\textwidth]{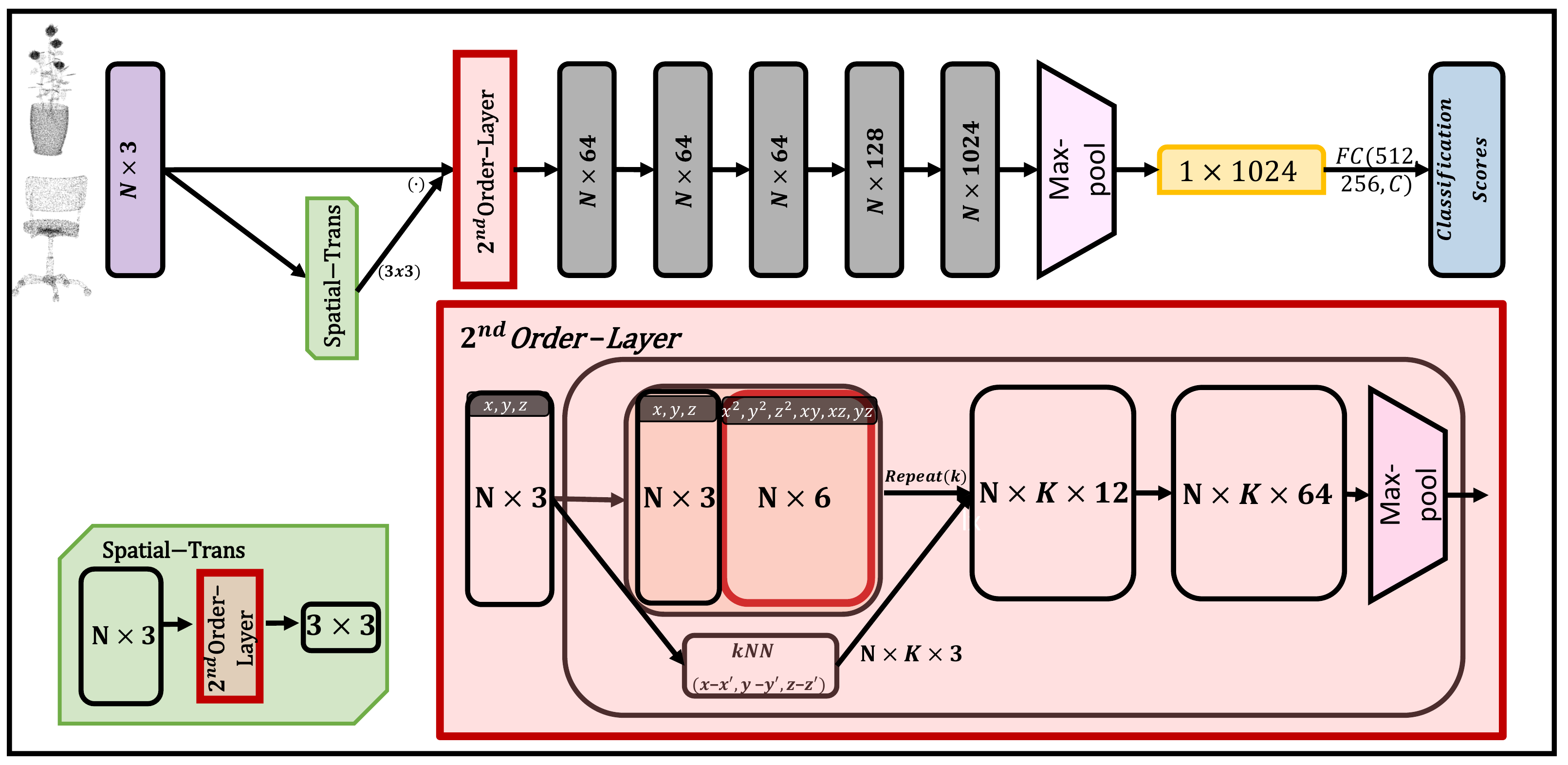}}
	\caption{\textbf{Momen$^e$t Architecture}. 
	Momen$^e$t takes as input $N$ points, applies a spatial transformation, and then aggregates the polynomial input multiplications with the k-nearest-neighbors graph as suggested in \cite{wang2018dynamic}. Next, point-wise MLP layers are followed by max-pooling. The output is a probability distribution over the $C$ classes. \textbf{$\boldsymbol{2^{nd}}$ Order Layer} aggregates each point separately with its second order polynomial expansions, then concatenates it with the point's neighbors.}
	\label{fig:comp}
\end{figure*}

The principal directions are defined by the eigenvectors of the second order moments matrix $\Sigma$, see Eq. \ref{eu_discreteMom}. 
They can be used to rotate and translate a given set of points into a canonical pose, where the axes 
 align with directions of maximal variations of the given point cloud  \cite{campbell1981geometry}. 
The first principal direction $d_1$, the eigenvector corresponding to the largest eigenvalue, 
is the axis along which the largest variance in the data is obtained.
For a set of points $X \subseteq \mathbb{R}^3$, the $k$th direction can be found by
\begin{equation} 
%\label{eu_TRANS}
d_k = \operatorname*{arg\,max}_{\|d\|=1} d^T 
    \left ( X_k X_k^T \right ) d,
\end{equation}
 where
\begin{equation} 
%\label{eu_TRANS}
X_k = X - \sum_{s=1}^{k-1}d_s d_s^T X. 
\end{equation}
\newline

%-------------------------------------------------------------------------

%___________________________________________

\subsection{Approximation of polynomial functions} 
In the suggested architecture we added low order polynomial functions as part of the input. 
The question arises whether a network can learn polynomial functions, obviating the need to add them manually. 
Here, we first provide experimental justification to the reason that one should take into account the ability of a network to learn such functions as a function of its complexity. 
Mathematically, we examined the ability of a network $\hat{f}(w,x)$ to approximate $f(x)$, where $w$ denotes the network parameters, such that 
\begin{equation} \label{eu_LINF}
\operatorname*{arg\,min}_{\hat{f}(w,x)}
\| \hat{f}(w,x)-f(x)\|_{L_{\infty}}<\epsilon,
\end{equation}
for a given function $f: \mathbb{R} \rightarrow \mathbb{R}$.
Theoretically, according to \cite{liang2016deep,yarotsky2017error},
there exists a ReLU network that can approximate polynomial functions up to the above accuracy on the interval $[0,1]$, with network depth, number of weights, and computation units of $\mathcal{O}(\epsilon^{-1})$ each.

In order to verify these theoretical claims, we performed experiments to check whether a network can learn the geometric moments from the point cloud coordinates.  
Figure \ref{fig:x2} shows an example for $f(x)=x^2$ and its approximation by a simple ReLU networks.

The pipeline can be described as follows.
First we consider uniform sampling of the interval of $[0,1]$; the number of samples was chosen experimentally to be $1,000$ samples. 
Next, we arbitrarily chose the number of nodes in each layer to be $4$, each with ReLU activation, using fully connected layers (in contrast to the suggested network where we perform MLP separately per point). 
Lastly, we set the weight initialization to be taken from a normal distribution. 

Ideally, networks with a larger number of layers could better approximate a given function. 
However, it is still a challenge to train such networks. 
Our experiments show that although a two layer network has achieved the theoretical bound, the network had difficulty to achieve more accurate results than $L_\infty = 0.006$ even when we increased the number of layers above 6. 
Furthermore, we tried to add skip connections from the input to each node in the network; however, we did not observe a significant improvement.

Comparing two point clouds by comparing their moments is a well known method in the geometry processing literature. 
Yet, we have just shown that the approximation of polynomial functions is not trivial for a network.
Therefore, adding polynomial functions of the coordinates as additional inputs could allow the network to learn the elements of $\Sigma$  in Eq. \ref{eu_discreteMom}
and, assuming consistent sampling, should better capture the geometric structure of the data.
\newline

%___________________________________
%\subsection{Learning Moments Attributes}
%We showed that the elements of second order moments are crucial to classify shapes, therefore we want to incorporate them as part of the feature vectors.
%Given a feature vector $A \in \mathbb{R}^{n}$ of one point we can represent $B \in \mathbb{R}^{f}$, the output feature vector, as:

%\begin{equation} 
%\label{eq_wxsq}
%B = (W_1 \cdot A, W_2\cdot vec(AA^T))
%\end{equation}

%Where $W_1 \in \mathbb{R}^{f \times n},W_2 \in \mathbb{R}^{f \times n^2}$ are learnable weighted matrices, $vec(C)$ stands for reshaping all elements of $C$ into a single column vector and (,) denotes concatenation of two vectors.

\subsection{Momen$^e$t Architecture}

The network architecture is as follows (figure \ref{fig:comp}): the network takes N points as input, applies a spatial transformation resulting in a 3x3 transformation matrix. We multiply each input point with the transformation matrix to align the point cloud. Then, the transformed point cloud is fed to our second-order layer, which is a simple polynomial function of the point cloud coordinates concatenated with the point's neighbors followed by an MLP layer. 
The higher order features are then processed by several point-wise MLP layers which map the points to higher dimensional space, followed by max-pooling. 
%Note that the third layer is a concatenation of the previous layer and the outputs of the 2nd order layers, while the input to the fifth layer is concatenation of all the previous layers. 
Finally, similarly to pointNet, we apply two fully connected layers of sizes $(512,256)$ and output a softmax over the classes.

The baseline architecture of the suggested Momen$^e$t network is based on the pointNet architecture. 
Our main contribution is the addition of polynomial functions as part of the input domain.
The addition of point coordinate powers to the input is a simple procedure that improves accuracy and decreases the run time during training and inference.
%To implement this, with previously $3$ elements for each point, and extended the input to $9$ elements.
We also calculated the k-Nearest-Neighbors (kNN) graph from the point coordinates as suggested in \cite{wang2018dynamic} with $k=20$. The output of the kNN block is the distance between a point to each of its $20$ neighbors, which is $(x-x',y-y',z-z')$ where $(x,y,z)$ and $(x',y',z')$ are the point and neighbor coordinates respectively.
Then, we use a tile operation on our 2nd order polynomial expansions with the kNN output, followed by a point-wise MLP layer.

\section{Experimental Analysis}

%We compared the performance of the proposed model to that of pointNet \cite{qi2017pointnet}, as their architectures are similar and operate directly on points in $\mathbb{R}^3$. 
%We used the pointNet implementation provided by the authors. 
%The following paragraphs describe the datasets and experimental results.

\subsection{Toy Problem}

\begin{figure}
\centering
\begin{subfigure}[b]{0.45\textwidth}
   \includegraphics[width=1\linewidth]{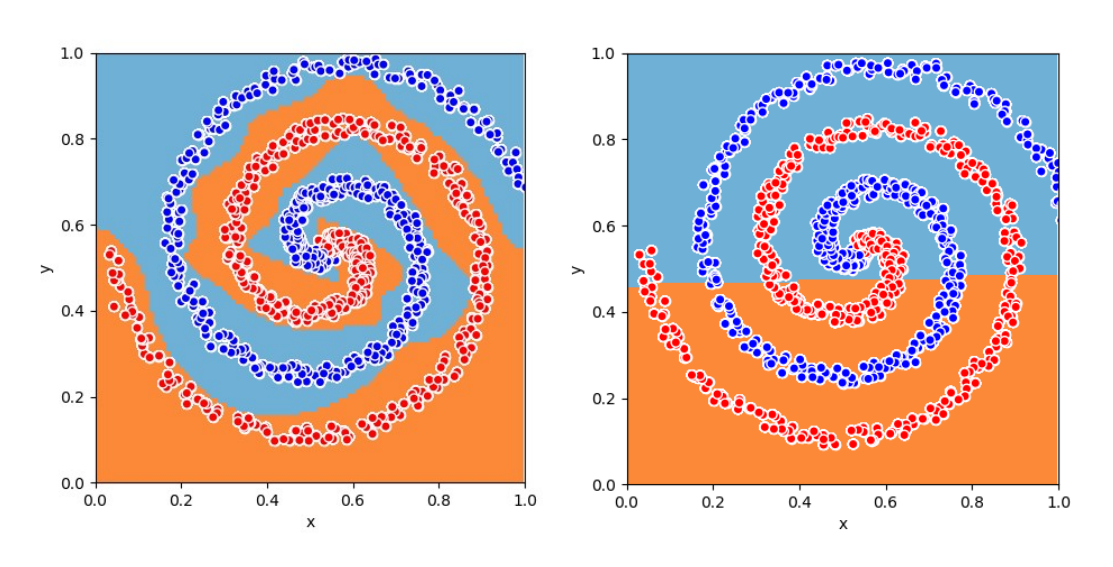}
   \begin{minipage}[t]{.5\linewidth}
\centering
\subcaption{}\label{fig:ours}
\end{minipage}%
\begin{minipage}[t]{.5\linewidth}
\centering
\subcaption{}\label{fig:regular}
\end{minipage}

\end{subfigure}
\begin{subfigure}[b]{0.45\textwidth}
   \includegraphics[width=1\linewidth]{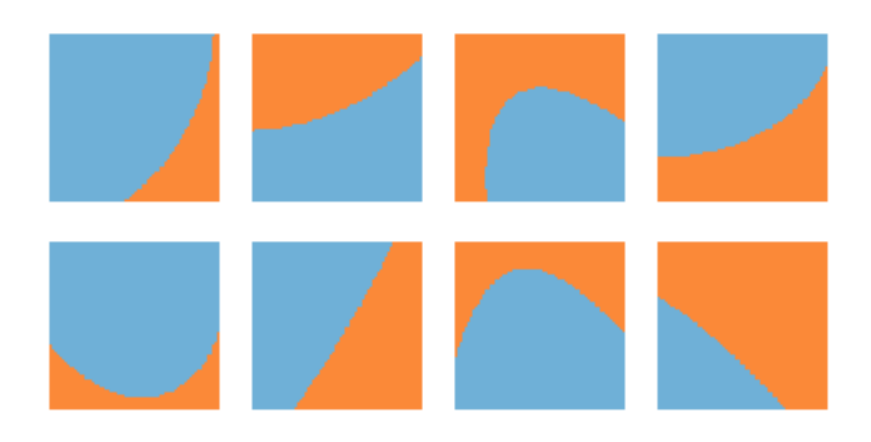}
   \caption{}
   \label{fig:boundries}
\end{subfigure}

\caption[2spiral]{Output response surfaces of a network comprised of 8 hidden nodes with polynomial expansions (a) and without (b). The output responses of each hidden node of (a) is presented in (c). The training set (data points) are colored in red/blue, which indicate positive/negative class respectively while orange/light blue areas are the corresponding network predictions.}
\label{fig:toyExample}
\end{figure}

%AZ: I like the toy example and the figure. Only problem - if someone prints a BW, the colors are all gray, and the whole discussion/argument is lost.   
We first present a toy example to illustrate that extra polynomial expansions as an input can capture geometric structures well.
%The two-spirals classification problem is a well-known benchmark that is challenging for most neural network architectures \cite{}. 
Figure \ref{fig:toyExample} (a+b) shows the data and the network predictions for $1000$ noisy 2D points, each point belongs to one of two spirals. 
We achieved $98\%$ success with extra polynomial multiplications, i.e $x^2,y^2,xy$, as input to a one layer network with only $8$ hidden ReLU nodes, contrary to the same network without the polynomial multiplications (only $53\%$ success).
%, which consist only 8 hidden ReLU nodes.
%With extra polynomial expansions, i.e $x^2,y^2,xy$, to the input we achieved 98\% success with only 8 hidden ReLU nodes.
Figure \ref{fig:boundries} shows the decision boundaries that were formed from the 8 hidden nodes of network \ref{fig:ours}. As expected, radial boundaries can be formed from the additional polynomial extensions and are crucial to forming the entire decision surface.

%Our version consist 2D point cloud with 1000 points with normal distribution noise ()
%Figure 2 (a+b) shows the data and the network prediction for the two-spirals problem, which consists of 1000 points, noise on the x-y-plane that belong to one of two spirals. 
%With the addition of $x^2,y^2,xy$ to the input we succeed to solve the problem with only 8 hidden ReLU nodes and compared to the results to only x-y coordinates as input (figure 5.b). 
%Figure 6 shows the decision boundaries that were formed from the 8 hidden nodes. As expected, radial boundaries can be formed from the additional polynomial extensions and are crucial to forming the entire decision surface.

%___________________________________
\subsection{Classification Performance}
%___________________________________
\textbf{Dataset and data processing.} Evaluation and comparison of the results to previous efforts is performed on the ModelNet40 benchmark \cite{wu20153d}. ModelNet40 is a synthetic dataset composed of Computer-Aided Design (CAD) models, containing $12,311$ CAD models given as triangular meshes, split to $9,843$ samples for training and $2,468$ for testing. Pre-processing of each triangular mesh as proposed in \cite{qi2017pointnet} yields $1024$ points sampled from each triangular mesh using the farthest point sampling (FPS) algorithm. Rotation by a random angle, about the $y$ axis, and additive noise are used for data augmentation. The database contains samples of very similar categories, like the flower-pot, plant and vase, for which separation is subjective rather than objective and is a challenge even for a human observer.

%_________________________

%\textbf{Training and test procedures.}  
%We use the preprocessing advised by the authors of \cite{qi2017pointnet}. We preprocessed each object by applying a 

%For comparison, we trained two pointNet versions as published by the authors. 
%The first is a vanilla version which we compare to our Momen$^e$t network (baseline). 
%The second version incorporates transformer blocks, and we compare it to Momen$^e$t with the transformer blocks (STN) as well. 
%We use the preprocessing advised by the authors.
%The idea of adding polynomial functions to the input domain is simple, induces low time and space complexity, and achieves better results compared to those realized by pointNet. 
%Training on ModelNet40 takes about $3$ hours to converge with Tensorflow \cite{abadi2016tensorflow} and Nvidia Titan X.

\textbf{Results.}
Table \ref{table:results_table_modelNet40} shows the results of the ModelNet40 classification task for various methods that assume different representations of the data.
%comparision to pointNet,Dgcnn
Comparing to DGCNN, our method achieves slightly better results, doing so while using kNN only on the input points, and not on the input to all layers as in DGCNN. When DGCNN is used similarly with respect to the kNN usage, they report a significant drop in performance ($91.9\%$), this highlights the power of geometric moments as features for point clouds.

%Table \ref{table:results_table} compares classification results when

%pointNet\cite{qi2017pointnet} and the suggested network Momen$^e$t are applied to ModelNet40 and S3DIS. 
%The results on  these two benchmark datasets confirm the superiority of the Momen$^e$t architecture.
%Other points based approaches \cite{simonovsky2017dynamic,wang2018dynamic,klokov2017escape,li2018pointcnn,qi2017pointnet++,ben20173d} report better results; however, they consider  features that require a support larger than a single point, or a partitioning of the input set of points in addition to the point features. 
% }
%It should be noted that although classification rates above 90\% were reported for example in  \cite{brock2016generative,maturana2015voxnet,qi2016volumetric}, they did not use point-clouds as input, but different data representations such as meshes, voxel-grids or multi-view images.

%One of our future research directions is to characterize and add local geometric features, in order to improve accuracy. 
%One should take into account that different data representations such as meshes, voxels, or multi-view contain more spatial information.

%We also tested the effects of input and feature transformations on the results. 
%Using spatial transformers improved the Momen$^e$t performance by 2.1\%. 
%We conclude that the suggested approach achieves substantially better results than pointNet, with or without the transformer blocks.

\begin{table}
\begin{center}
\begin{tabular}{l  c c c } 
 \hline
Method & \shortstack{Mean Class\\ Accuracy}  & \shortstack{Overall\\ Accuracy} \\ 
%\shortstack{Mean Class \\ Accuracy}  & \shortstack{Overall \\  Accuracy} \\ 
  
   \hline  
  PointNet \cite{qi2017pointnet}  & 86.2 &89.2\\ 
  Deep-Sets \cite{zaheer2017deep} & - & 87.1\\
  PointNet++ \cite{qi2017pointnet++}  & - &90.7\\
  PointCNN \cite{li2018pointcnn} &  - &91.7\\
  ECC \cite{simonovsky2017dynamic}&83.2 & 87.4\\
  DGCNN \cite{wang2018dynamic} & 90.2 & 92.2 \\
  Kd-Networks \cite{klokov2017escape} &88.5 &91.8\\
  SO-Net \cite{li2018so}               & 87.3 & 90.9\\
 KC-Net    \cite{shen2018mining}     & - & 91.0\\
 ShapeContexNet \cite{xie2018attentional}  & 87.6&90.0 \\
 PCNN   \cite{Atzmon:2018:PCN:3197517.3201301}        & - & 92.3\\
 3DmFV-Net   \cite{ben20183dmfv}         & - & 91.4 \\
 \hline  
 \hline 
 Momen$^e$t   & \textbf{90.3} & \textbf{92.4}\\ 
 \hline
\end{tabular}
\caption{Comparison of classification accuracy (\%) on ModelNet40. }
\label{table:results_table_modelNet40}
\end{center}
\end{table}

%\begin{table}[!ht]
%\begin{center}
%\begin{tabular}{ c|c|c } 
% \hline
%Method & \quad S3DIS \quad & \quad ModelNet40 \quad \\ 
%  \hline
%  PointNet (Baseline)& 65.0 &86.6\\ 
%  PointNet  & 66.4 &88.9\\ 
% \hline 
% Momen$^e$t (Baseline)& 66.1 &87.2\\ 
% Momen$^e$t  & 66.7 &89.3\\ 
% \hline
%\end{tabular}
%\caption{Comparison of classification accuracy (\%) on ModelNet40 and S3DIS datasets.}
%\label{table:results_table}
%\end{center}
%\end{table}

\subsection{Ablation Experiments} % More Experiments?

\begin{figure}[htbp]
	\centering
	\centerline{\includegraphics[width=0.9\linewidth]{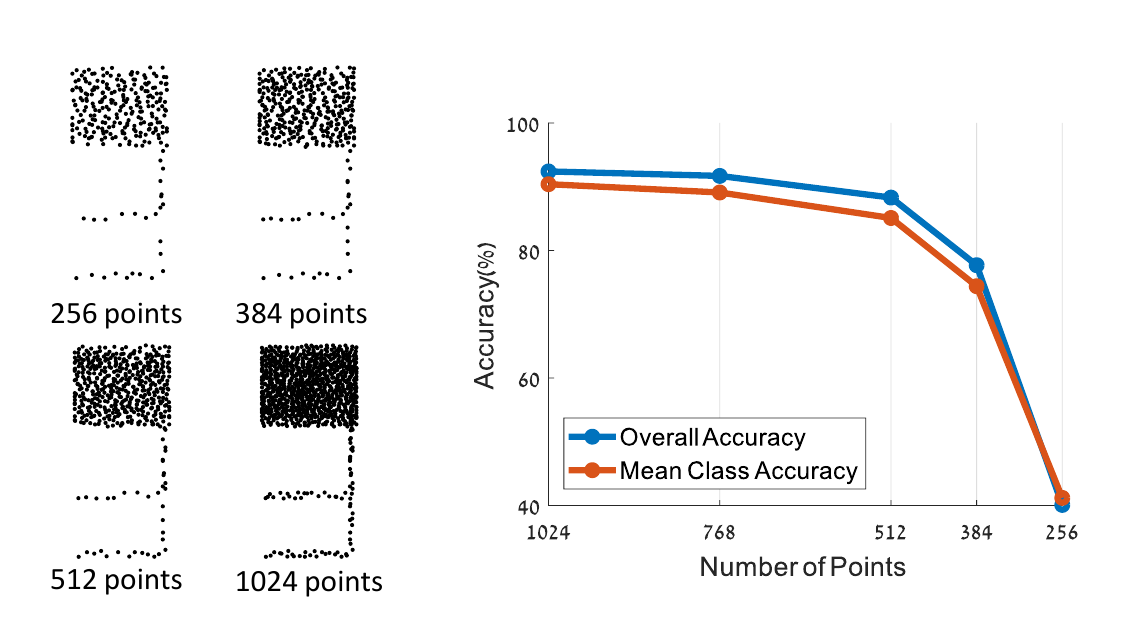}}
	\caption{Left: Point clouds with sparse to dense sampling.  Right: Results of our model tested with random input dropout. The model is trained with 1024 points.}
	%AZ consider: "Point clouds with sparse to dense sampling"..."  MJ: ok
	\label{fig:diff_num}
\end{figure}

\textbf{Robustness to point density and different orientations.}
We evaluate the robustness of our model to variations in point cloud density and to different orientations. For these tests, the model was trained in the same manner as we reported in previous sections, but was evaluated with the augmented data.
Figure \ref{fig:diff_num} shows results for the point density test. The model is fairly robust to sampling density variation when the ratio of dropped points is less than half. 

\begin{figure}[htbp]
	\centering
	\centerline{\includegraphics[width=1\linewidth]{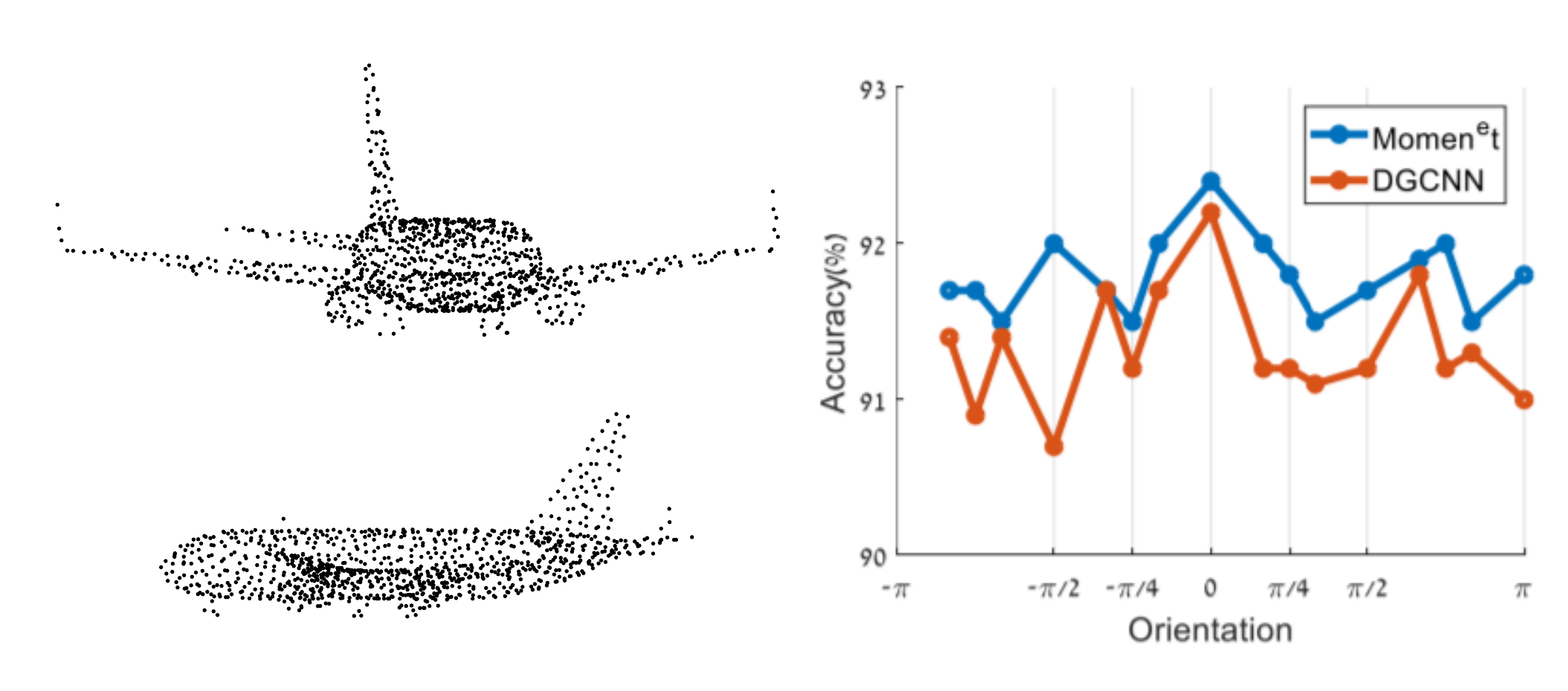}}
	\caption{Left: Point clouds in different orientations.  Right: Comparison of our model and DGCNN for different orientations along y-axis. The model is trained with random rotations along y-axis as DGCNN.}
	\label{fig:orien}
\end{figure}

Next, we simulate different rotations of the point clouds along the y-axis. Figure \ref{fig:orien} shows a comparison between our method and DGCNN, our model is more robust for almost all rotations when compared to DGCNN.
\newline

\textbf{Variations of the Architecture.}
To verify the effectiveness of adding the $2^{nd}$-order layer, we test different variations in model architecture with respect to the ModelNet40 dataset. 
First, we examine different polynomial orders of the point cloud coordinates, see table \ref{table:order_variation}. It can be seen that adding polynomial functions of order higher than $2$ does not yield improvements in performance. 

\begin{table}[h]
\begin{center}
\begin{tabular}{l c  c  c  c}
 \hline
\quad Order \quad \quad &  \shortstack{Overall Accuracy} \\
\hline
$2^{nd}$   &   92.4\\
$2^{nd}+3^{rd}$     & 92.3\\
$3^{rd}$  &  91.9 \\
Learnable & 91.6 \\
\hline
\end{tabular}
\end{center}
\caption{Overall accuracy of modelNet40 for different order of polynomial expansions. }
\label{table:order_variation}
\end{table}

We also tested a variation in which the order of polynomial functions was learned by a special neural network layer defined as $exp(w \cdot log(|x|+\epsilon))$ where $w$ is the learnable kernel and $x$ is the input \cite{trask2018neural}.
Second, we tested the effect of different components in our network, see table \ref{table:variation}. Similar to our previous tests, removing the spatial transformer and reducing the number of layers in our network still achieves better accuracy than the baseline version of DGCNN ($91.2\%$).
\newline

\begin{table}[h]
\begin{center}
\begin{adjustbox}{width=1\linewidth}
\begin{tabular}{c  c  c  c}
 \hline
 \shortstack{Spatial\\Transformation} & Layers & kNN & \shortstack{Overall \\Accuracy} \\
\hline
\xmark  & [64,64,64,128,1024] & \xmark & 88.1 \\
\xmark &[256,1024] & \cmark & 91.4\\
\xmark  & [64,64,64,128,1024] & \cmark & 92.1 \\
\cmark & [64,64,64,128,1024] & \cmark &  92.4\\
\end{tabular}
\end{adjustbox}
\end{center}
\caption{Variation of the network architecture. The effect of different components in our network on the accuracy with respect to ModelNet40 dataset.}
\label{table:variation}
\end{table}

\begin{table}[h]
\begin{center}
\begin{adjustbox}{width=\linewidth}
\begin{tabular}{l c  c  c  c}
 \hline
\quad Method \quad \quad &  Memory (MB) &   \shortstack{Inference\\Time (msec)}  & Acc \\
\hline
%PointNet (Baseline)  & 9.4& 11.6& 87.1\\
PointNet \cite{qi2017pointnet} & 40 & 5.6 & 89.2\\
PointNet++ \cite{qi2017pointnet++} & 12 & 10.4 & 90.7 \\
%DGCNN (Basline) & 11 & 29.7 & 91.2 \\
DGCNN \cite{wang2018dynamic}& 21 & 17.3 &92.2\\
PCNN \cite{Atzmon:2018:PCN:3197517.3201301} & 17 & 54.1 & 92.3 \\
%PointNet++ \cite{qi2017pointnet++}& 0.8M & 150M\\
\hline
\hline
%Momen$^e$t (Baseline) &  &  & \\
Momen$^e$t &20 & 9.6 & 92.4\\

\end{tabular}
\end{adjustbox}
\end{center}
\caption{Comparison of time and space complexity.
Memory is the size of the model in mega-byte (MB) and the inference run-time was measured in milli-seconds (msec). Acc stands for the suggested network accuracy on ModelNet40. 
%Here, M stands for $10^6$. 
%The Momen$^e$t implementation achieved a reduction of FLOPs by more then 90\% and of model size by 20\% compared to pointNet architecture.
}
\label{table_FLOPS}
\end{table}

\begin{table*}[h]
\begin{center}
%\begin{tabular}{l  c | c | c c c c c c c c c c c c c c c c} 
\begin{adjustbox}{width=\textwidth}
\begin{tabular}{l  c | c c c c c c c c c c c c c c c c c}
 \hline
 & Mean & Air. & Bag & Cap & Car & Chair & Ear. & Guitar & Knife & Lamp &Laptop & Motor & Mug & Pistol & Rocket & Skate. & Table \\ 
%\shortstack{Mean Class \\ Accuracy}  & \shortstack{Overall \\  Accuracy} \\ 
  
   \hline  
  PointNet \cite{qi2017pointnet}& 83.7 &83.4& 78.7& 82.5& 74.9& 89.6& 73& \textbf{91.5}& 85.9& 80.8& 95.3& 65.2& 93& 81.2& 57.9& 72.8& 80.6 \\ 
  DGCNN \cite{wang2018dynamic} &\textbf{85.1}&84.2&\textbf{83.7}&84.4&77.1&\textbf{90.9}&\textbf{78.5}&\textbf{91.5}&\textbf{87.3}&\textbf{82.9}&96.0&67.8&93.3&82.6&59.7&\textbf{75.5}&82.0 \\
  Kd-Net \cite{klokov2017escape}& 82.3 & 80.1 & 74.6 & 74.3 & 70.3 & 88.6 & 73.5 & 90.2 & 87.2 & 81.0 & 94.9 & 57.4 & 86.7 & 78.1 & 51.8 & 69.9 & 80.3\\
  SO-Net \cite{li2018so}   & 84.9 & 82.8 & 77.8 & 88.0 & 77.3 & 90.6 & 73.5 & 90.7 & 83.9 & 82.8 & 94.8 &69.1 & 94.2 & 80.9 & 53.1 & 72.9& \textbf{83.0 } \\
 KC-Net   \cite{shen2018mining}&  83.7& 82.8& 81.5& 86.4& 77.6& 90.3& 76.8& 91.0 &87.2& 84.5& 95.5 &\textbf{69.2} &94.4& 81.6&\textbf{ 60.1}& 75.2& 81.3   \\
 PCNN \cite{Atzmon:2018:PCN:3197517.3201301} &\textbf{85.1}& 82.4& 80.1& 85.5&\textbf{ 79.5}& 90.8& 73.2& 91.3& 86.0& 85.0& 95.7& 73.2& 94.8& \textbf{83.3}& 51.0& 75.0& 81.8\\
 3DmFV-Net \cite{ben20183dmfv} &84.3& 82.0& 84.3& 86.0& 76.9& 89.9& 73.9& 90.8& 85.7& 82.6& 95.2& 66.0& 94.0& 82.6& 51.5& 73.5& 81.8 \\
 \hline  
 \hline 
 Momen$^e$t   & 84.9 & \textbf{ 84.4} & 82.7 &\textbf{88.4} & 78.5 &90.7 & 78.1 &90.1 &\textbf{ 87.3} & 82.5 &\textbf{ 96.1} &65.6&\textbf{94.9}&83.1&58.5&73.6&81.7\\ 

\end{tabular}
\end{adjustbox}
\caption{Comparison of segmentation results on the ShapeNet database. The evaluation metric is mean Intersection over Union (IoU). }
\label{table:results_table_shapenet}
\end{center}
\end{table*}

\textbf{Memory and Computational Efficiency.} 
%As a result of different representations such as meshes, multi-view and voxel-grids, 3D DNN architectures present a wide range of computational requirements. 
%For example, the MVCNN DNN architecture \cite{su2015multi} performs $62$ billion floating-point arithmetic operations to classify a point cloud.
Our network is implemented in tensorflow \cite{abadi2016tensorflow}, we use ADAM optimizer \cite{kingma2014adam} with initial learning rate 0.01. 
Table \ref{table_FLOPS} presents computational requirements with respect to the number of the network's parameters (memory) and with respect to the inference time required by the models.

The inference time is measured for one point cloud with 1024 points on nVidia Titan X GPU.
In comparison with other methods, our results show that adding polynomial expansions to the input leads to better classification performance, as well as computational and memory efficiency.

\subsection{Part Segmentation Performance}
\textbf{Dataset  and  data  processing.}
Evaluation and comparison of the results to previous efforts is performed on the ShapeNet benchmark \cite{wu20153d}. ShapeNet contains $16,881$ 3D models with $50$ annotated parts from $16$ categories. The dataset is considered challenging due to the imbalance of its labels.
The aim of the part segmentation task is to predict a category label to each point in the point cloud. 
The splitting of the dataset to train/test and sampling of the points were done similarly to \cite{qi2017pointnet}.

\textbf{Training and test procedures.}
We extended our architecture to address the segmentation task, which requires local and global features. 
The network segmentation architecture is as follows: 
the network takes N points as input, applies a spatial transformation, then the transformed point cloud is fed to our second-order layer as in the classification network. 
The higher order point-wise features are processed by point-wise MLP layers and their outputs are concatenated to the $1024$ global features layer. Finally, we apply several point-wise MLP layers to produce a label to each point.

\begin{figure}[htbp]
	\centering
	\centerline{\includegraphics[width=1\linewidth]{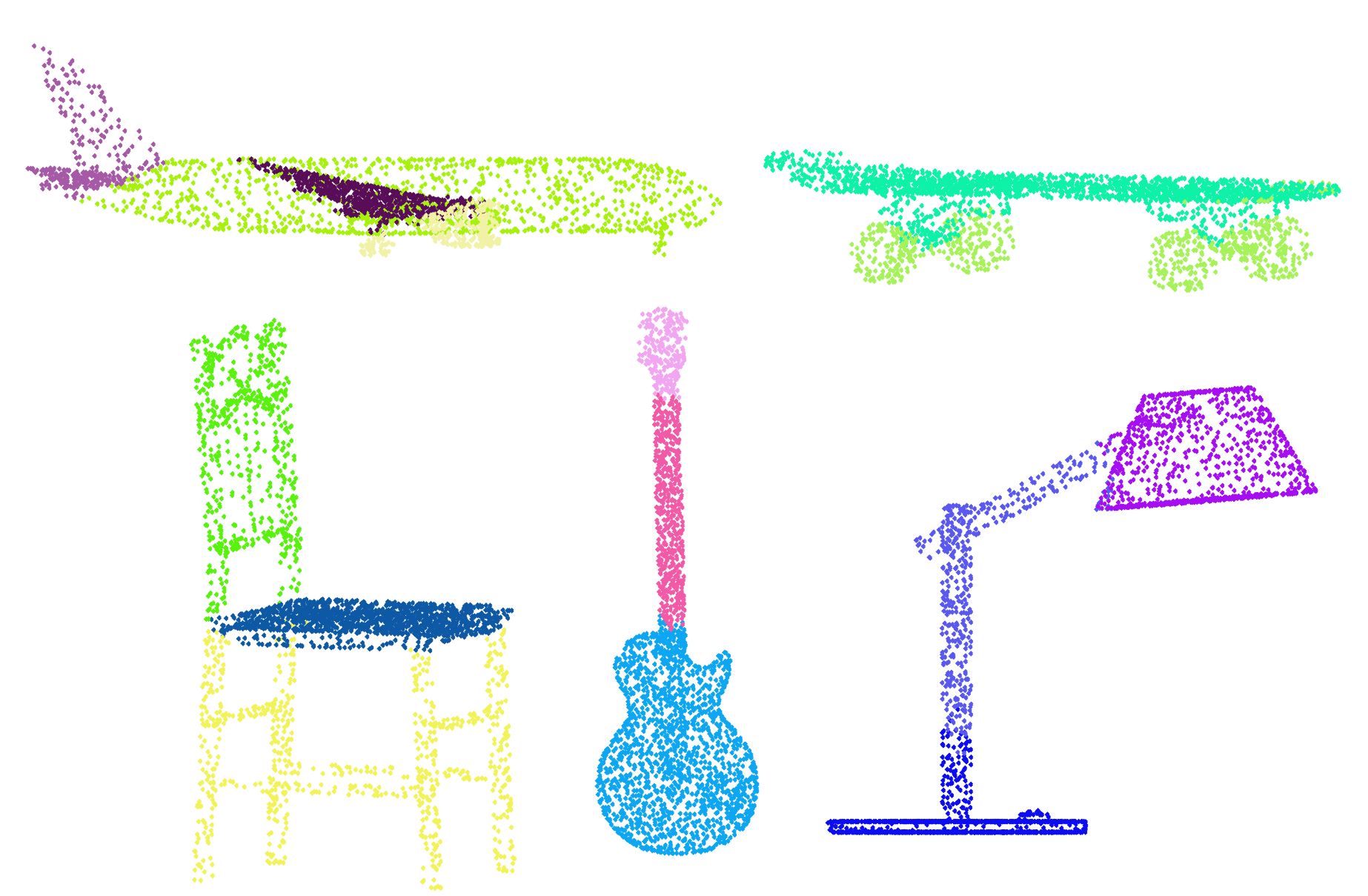}}
	\caption{Qualitative results of Momen$^e$t on the part segmentation task, performed on the ShapeNet database \cite{wu20153d}.}
	\label{fig:qualitive}
\end{figure}

To reduce memory consumption, we decreased the k-nearest neighbors to only $15$, compared to $k=30$ in DGCNN. 
Decreasing the number of neighbors during the kNN computation also helps to substantially reduce the run-time of the network during both training and inference.

%The evaluation metric is mean Intersection over Union (IoU). 

\textbf{Results.}
The evaluation metric is mean Intersection over Union (IoU), and in order to calculate it we followed the same scheme as pointNet. 
We compute the IoU of each category by averaging the IoU of all the samples belonging to that category, and then average the IoU of all the categories. 
Lastly, we average the IoU of all the test samples to get our final performance metric.
Table \ref{table:results_table_shapenet} shows detailed segmentation results on the ShapeNet dataset. 
Our network achieves similar results to state-of-the-art methods despite the fact that we use only $15$ neighbors, which leads to a reduction in memory consumption and run-time. 
Qualitative results are shown in Figure \ref{fig:qualitive}.

\section{Conclusions}

In this paper, we combined a geometric understanding about the ingredients required to construct compact shape signatures with neural networks that operate on clouds of points to leverage the network's abilities to cope with the problem of rigid objects classification. 
By lifting the shape coordinates into a small dimensional, moments-friendly space, the suggested network, Momen$^e$t, is able to learn more efficiently in terms memory and computational complexity and provide more accurate classifications compared to related methods.
Experimental results on two benchmark datasets confirm the benefits of such a design.
We demonstrated that lifting the input coordinates of points in $\mathbb{R}^3$ into $\mathbb{R}^9$ by simple second degree polynomial expansion, allowed the network to lock onto the required moments and classify the objects with better efficiency and accuracy compared to previous methods that operate in the same domain.
We showed experimentally that it is beneficial to add these expansions for the classification and the segmentation tasks.
We believe that the ideas introduced in this paper could be applied in other fields where geometry analysis is involved, and that the simple cross product of the input point with itself could improve networks abilities to efficiently and accurately handle geometric structures. 

\section*{Acknowledgments}
This research was partially supported by Rafael Advanced Defense Systems Ltd.
%AZ: homogeneous coordinates ?? I think its Cartesian coordinates. or just say "cross product of the input point with itself..." MJ:fixed

%AZ: maybe change all "2nd" to $2^{nd}$ or $\nth{2}$ MJ:thanks!

%AZ: modelnet40, Shapenet citations in captions of figure/tables -  W or WO, just be consistent. MJ: fixed

{\small
\bibliographystyle{unsrt}
\bibliography{main}
}

\end{document}